\pdfoutput=1

\documentclass{ithaca}

\NUMERO{MCXIV}
\ANNO{2067}

\begin{document}
\title{L'universo emergente della gravit\`a quantistica}

\dottacitazione{Possiamo davvero \lq\lq conoscere" l'universo? Mio Dio, è già così difficile orientarsi a Chinatown. - Woody Allen}


\author{
\name{Daniele Oriti}
\affiliation{Max Planck Institute for Gravitational Physics (Albert Einstein Institute) \\ Am M\"uhlenberg 1, D-14476, Potsdam, Germany, EU \\ Arnold-Sommerfeld-Center for Theoretical Physics, Ludwig-Maximilians-Universit\"at \\ Theresienstrasse 37, D-80333 M\"unchen, Germany, EU}
}
\ithacamaketitle
\abstract{Una introduzione al problema della gravit\`a quantistica, alle recenti prospettive su uno spaziotempo emergente, e alla loro realizzazione (potenziale) nel contesto delle teorie di campo su gruppi, in cui l'universo emerge come un condensato di costituenti elementari non-spaziotemporali. }

\section{Il problema della gravit\`a quantistica}
Chiariamo innanzitutto di \textit{cosa} stiamo parlando: cosa studia la gravit\`a quantistica?

Guardatevi attorno. Vedrete tanti corpi materiali, fatti di atomi, in movimento e in interazione l'uno con l'altro, tramite forze elettriche o nucleari (le particelle che compongono gli atomi cos\`i interagiscono). Gli stessi corpi cadono a terra oppure orbitano uno attorno all'altro (nel caso in cui siate all'aperto sotto un cielo stellato e non seduti al chiuso di una stanza). In quest'ultimo caso, stiamo assistendo alla loro interazione gravitazionale. Chi lavora in gravit\`a quantistica studia questa interazione.

Ma questo non rende ancora l'idea. Tornate a guardarvi attorno, adesso facendo uso di un po' di immaginazione (e ce ne vuole parecchia per fare gravit\`a quantistica). Immaginate di rimuovere uno ad uno tutti i corpi che vedete attorno a voi, facendo finta che non ci siano. Cosa rimane, quando avete finito? Nulla, direte voi. E lo stesso pensavamo in molti, fino all'inizio del secolo scorso. Oppure direte: rimane lo spazio vuoto, dentro cui si muovono i corpi materiali, che per\`o non \`e altro che un contenitore inerte e vuoto, appunto, niente di troppo interessante. Questo \`e quello che pensavano tutti gli altri, inclusi molti fisici. Ecco questo \`e quello che, pi\`u esattamente, studia chi si occupa di gravit\`a quantistica. Non molto eccitante, detta cos\`i.
Ci\`o che lo rende eccitante \`e quello che abbiamo imparato dalla Relativit\`a Generale, un secolo fa.

La lezione principale di questa teoria, bella come poche altre, \`e che la gravit\`a non \`e altro che la geometria dello spazio, o meglio dello spaziotempo, stesso (Figura ~\ref{spacephysical}).

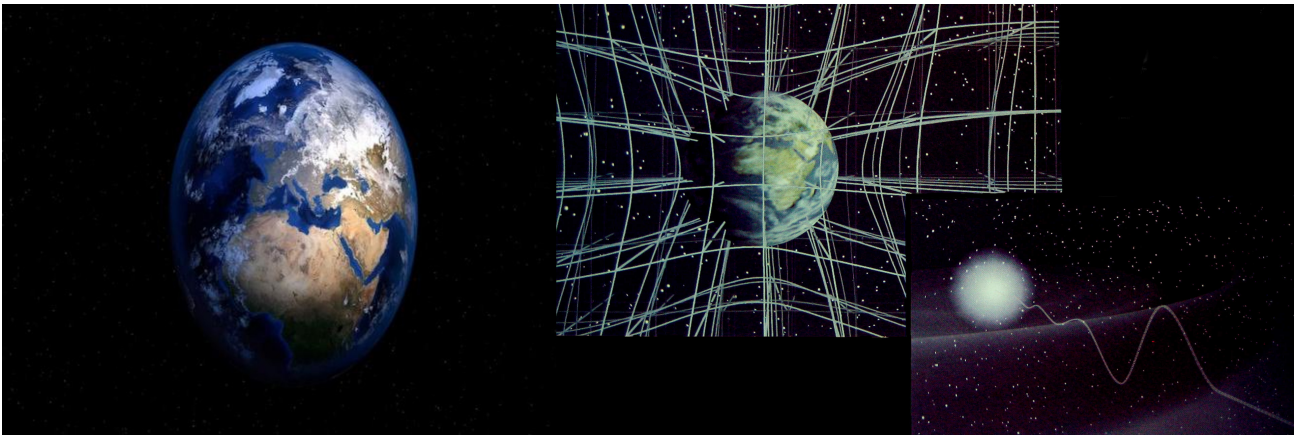
\WideFigure{space-physical-GR}{\label{spacephysical}Lo spazio(tempo) stesso diventa un sistema fisico e dinamico, in Relativit\`a Generale}

Pi\`u precisamente, quando diciamo che due corpi materiali si attraggono tramite interazione gravitazionale stiamo in realt\`a dicendo che questi due corpi deformano lo spaziotempo intorno ad essi, in una maniera  dipendente dalla loro massa e dalla loro energia (inclusa quella dovuta al loro stesso movimento, quella dovuta alla loro carica elettrica, ecc.), e che si muovono liberamente (cio\`e senza che ci sia ad agire su di essi alcuna \lq\lq forza\rq\rq) in questo spaziotempo deformato. La deformazione a cui ci riferiamo \`e quella codificata da tutte le misurazioni di distanze spaziali e temporali tra oggetti materiali ed eventi fisici, di angoli e volumi, cio\`e appunto della \textit{geometria} dello spaziotempo. In un certo senso questi aspetti geometrici sono la \textit{definizione} di spaziotempo. Ecco la Relativit\`a Generale ci ha insegnato che lo spaziotempo stesso \`e un sistema fisico in  s\`e, che interagisce con gli altri sistemi fisici (i campi materiali, gli altri campi di interazione come quello elettromagnetico, ecc.), che ha delle propriet\`a intrinseche e una propria dinamica, per descrivere le quali essa ci fornisce le equazioni appropriate. Per quanto possa sembrare sorprendente, lo spaziotempo stesso diventa un attore alla pari degli altri campi di interazione e dei corpi materiali, piuttosto che il palcoscenico inerte, della grande rappresentazione teatrale cosmica.

Non \`e stato e non \`e facile interiorizzare questa lezione, cos\`i controintuitiva. Ma ormai la Relativit\`a Generale \`e la base di tutta l'astrofisica, di tutta la cosmologia (la scienza che studia come \`e nato e come evolve l'universo nel suo insieme), e ha ricevuto una quantit\`a enorme di conferme osservative. L'ultima, da mozzare il fiato, la rilevazione diretta delle onde gravitazionali, solo tre anni fa. E le onde gravitazionali sono niente altro che piccole increspature, deformazioni in movimento, dello spaziotempo stesso, analoghe alle increspature che si propagano sulla superficie di un fluido materiale. Una specie esotica di fluido, questa \`e infatti un'altra maniera intuitiva di pensare allo spaziotempo relativistico, come vedremo meglio nel seguito.

Ok, ma questo \`e quello che studia gi\`a la Relativit\`a Generale. Dove e perch\'e serve la gravit\`a quantistica? Il fatto \`e che sappiamo ormai che, nonostante tutti i suoi successi, la Relativit\`a Generale non \`e abbastanza. Non lo \`e innanzitutto perch\`e \textit{tutti gli altri sistemi fisici} sono molto ben descritti da un formalismo totalmente diverso, la Meccanica Quantistica (pi\`u precisamente la Teoria Quantistica dei Campi), basata su una matematica diversa e, soprattutto, un apparato concettuale che contraddice la gran parte dei pilastri su cui si basa la Relativit\`a Generale. In particolare, la Meccanica Quantistica tratta lo spaziotempo alla maniera di Newton, come uno sfondo inerte piuttosto che come un sistema fisico esso stesso. Non ha assorbito, in altre parole, la lezione principale della Relativit\`a Generale, la quale d'altra parte non ha assorbito nessuna delle lezioni che la fisica quantistica ci ha impartito riguardo il comportamento dei sistemi fisici. Viviamo quindi, come fisici, un mondo schizofrenico, in cui siamo costretti ad utilizzare due apparati concettuali incompatibili l'uno con l'altro, a seconda che si stia cercando di dar conto dei fenomeni gravitazionali, cio\`e dello spaziotempo in  s\`e, o del comportamento preciso (quindi quantistico) degli altri sistemi fisici.
Abbiamo ovviamente una quantit\`a di modelli semplificati, e di approssimazioni utili, che ci permettono di fare fisica nonostante questa incompatibilit\`a di fondo. Ma hanno validit\`a limitata e non danno quindi tutte le risposte che vorremmo avere, e soprattutto si basano su assunzioni e ipotesi, che solo una teoria pi\`u fondamentale  pu\`o giustificare o modificare.
\`E questa teoria pi\`u fondamentale, questa base pi\`u solida e completa (concettualmente quanto fisicamente) che ci manca per capire il mondo, \`e la gravit\`a quantistica \cite{isham, carlo, carlip}.

In sintesi, construire una teoria di gravit\`a quantistica vuol dire ottenere una comprensione pi\`u completa e pi\`u profonda della natura di spazio e tempo, e della loro interazione con la materia. E se lo spaziotempo in  s\`e \`e l'oggetto di studio, dobbiamo ottenere una descrizione del mondo in cui lo spaziotempo non \`e un dato di partenza ma qualcosa da capire. Dobbiamo pensare il mondo \textit{senza ipotizzare l'esistenza dello spaziotempo}.

\section{Quale fisica}
Messa cos\`i, potrebbe sembrare una questione puramente concettuale, da lasciare ai filosofi. Non \`e cos\`i.
Il punto generale \`e, infatti, che non abbiamo una teoria consistente e completa per trattare l'interfaccia tra fisica gravitazionale e fisica quantistica, cio\`e tutte quelle situazioni fisiche in cui sia le propriet\`a quantistiche dei corpi materiali e delle loro interazioni sia i loro effetti gravitazionali e la dinamica propria dello spaziotempo sono rilevanti. Per esempio, queste situazioni sono le fasi iniziali dell'universo (quindi la cosmologia primordiale) e i buchi neri (quindi una parte della astrofisica relativistica).

Enunciata la questione generale, cerchiamo di chiarire meglio cosa rende la gravit\`a quantistica un problema \textit{fisico} importante.

Chiediamoci innanzitutto a che scale di distanze ed energie dovrebbe essere rilevante. Se devono essere rilevanti sia gli effetti relativistici, sia quelli quantistici, sia quelli gravitazionali, in qualunque formulazione di una teoria di gravit\`a quantistica devono comparire la velocit\`a della luce $c$, la costante di Planck $\hbar$, e la costante gravitazionale di Newton $G$. Una loro combinazione d\`a una misura di lunghezza: $l_p = \sqrt{\frac{G\hbar}{c^3}} \approx 10^{-33}\,$ cm, di tempo $t_p = \sqrt{\frac{G\hbar}{c^5}}$, di energia: $E_p = \sqrt{\frac{c^5\hbar}{G}}  \approx 10^{19}\,$ GeV, e di curvatura $R_p \approx \frac{1}{l_p} \approx 10^{33}$ cm$^{-1}$, che definiscono la cosiddetta \lq scala di Planck\rq ~(vedi Figura \ref{planckscale}).

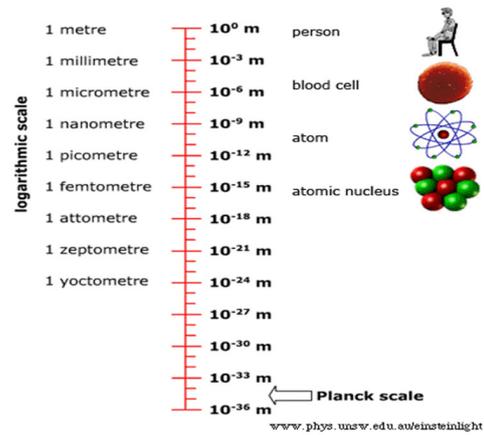
\ColumnFigure{PlanckScale}{\label{planckscale}La lunghezza di Planck, in confronto con altre scale di lunghezza pi\`u familiari.}

\`E evidente che stiamo parlando di scale molto distanti da qualunque fenomeno riproducibile in laboratorio o negli acceleratori (l'LHC raggiunge energe dell'ordine del TeV ($\approx 10^3$ GeV): tra la scala di Planck e le distanze pi\`u piccole mai testate, quelle dei quark, c'\`e circa a stessa distanza che tra i quark stessi e noi umani! Insomma, la  gravit\`a quantistica governerebbe solo fenomeni a distanze piccolissime ed energie enormi, e rimpiazzerebbe la Relativit\`a Generale solo quando la curvatura dello spaziotempo diventa parimenti enorme. Una conseguenza immediata \`e che non ci possiamo aspettare osservazioni dirette di tali fenomeni, e quindi che la costruzione di una teoria di gravit\`a quantistica venga guidata direttamente da input sperimentali.

Questo rende le cose molto pi\`u difficili, e storicamente inusuali per un fisico teorico. Vuol dire anche che la gravit\`a quantistica \`e destinata a rimanere speculazione o gioco matematico, o che solo alla matematica (e magari a criteri estetici) possiamo affidarci per giudicare la validit\`a delle verie proposte di teoria? Assolutamente no!

Per quattro ragioni.

Intanto, anche se le modifiche alla fisica che conosciamo, indotte dalla gravit\`a quantistica, fossero dell'ordine della scala di Planck, potrebbero bene esistere meccanismi fisici di amplificazione tali da renderle osservabili. Un esempio \`e quello di modifiche alla propagazione della luce o delle particelle materiali che portino a discrepanze (rispetto alla teoria dei campi usuale) che si accumulino con la distanza percorsa. Immaginate due particelle identiche emesse contemporaneamente da una galassia a distanza cosmologica da noi, e nella nostra direzione, e supponiamo che gli effetti di gravit\`a quantistica causino la loro velocit\`a relativa essere diversa (contrariamente a quello che risulterebbe  dalla Relativit\`a Generale). Il loro tempo di arrivo sui nostri telescopi sarebbe di conseguenza leggermente diverso. Quanto? dipende da quanto tragitto hanno percorso; se pure la differenza di velocit\`a/energia  di una rispetto all'altra fosse piccolissimo, dell'ordine della scala di Planck (cio\`e $\approx 1/E_p$), se il tempo trascorso prima di finire sui nostri telescopi fosse enorme, il ritardo accumulato potrebbe essere grande abbastanza da essere osservabile.
Un'altra possibilit\`a \`e che effetti di gravit\`a quantistica portino a violazioni di simmetrie fondamentali (per esempio quelle alla base della stessa Relativit\`a Generale). In questo caso, fenomeni che sarebbero semplicemente proibiti sulla base delle teorie usuali, diventerebbero possibili ed esperimenti di precisione che li rivelassero (non importa quanto raramente o debolmente manifesti) diventerebbero importanti input osservativi nella costruzione della teoria. Queste due classi di possibilit\`a sono infatti la base di una vasta letteratura riguardo la possibile fenomenologia della gravit\`a quantistica.

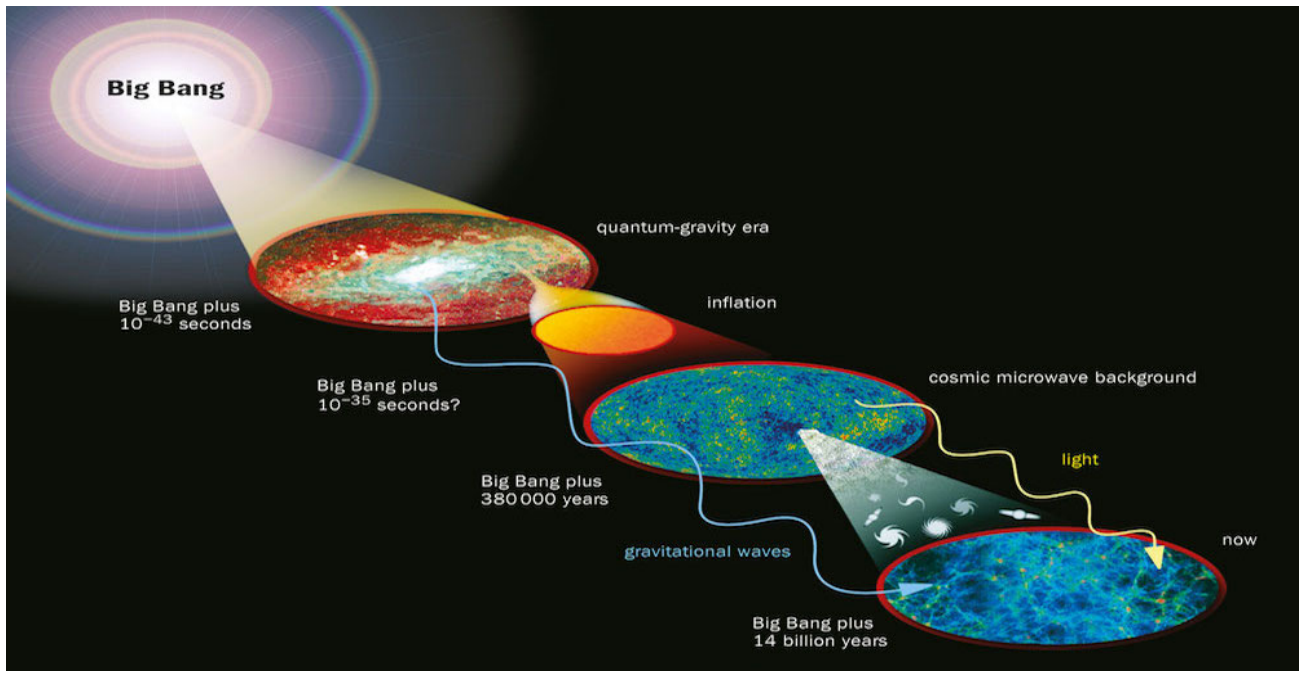
\WideFigure{cosmic-evolution}{\label{evolution} Le fasi chiave dell'evoluzione dell'universo, compresa (oltre alla fase di inflazione, che \`e ben accreditata, ma non senza alternative plausibili) quella di cui non sappiamo nulla e per descrivere la quale serve una teoria di gravit\`a quantistica: il big bang. (Immagine prodotta da Dana Baram - https://www.pinterest.de/danabaram1/)}

Un'altra ragione per considerare la gravit\`a quantistica a tutti gli effetti una questione fisica \`e che curvature spaziotemporali grandi abbastanza da chiamare in causa aspetti quantistici del campo gravitazionale vengono prodotte all'interno dei buchi neri e nelle fasi iniziali della vita dell'universo, al \lq big bang\rq ~(Figura \ref{evolution}).
In entrambi i casi, non abbiamo una teoria completa per descrivere cosa succede in queste circostanze. Sappiamo anche che descrizioni alternative di  ci\`o che succede (suggerite da modelli diversi di  gravit\`a quantistica) hanno conseguenze rilevanti dal punto di vista osservativo.

Facciamo un esempio. La radiazione di fondo cosmica, cio\`e la prima luce che ci giunge dopo essere stata emessa poco tempo ($\approx 10^5$ anni) dopo il big bang, \`e la base della cosmologia osservativa e della nostra comprensione dell'universo primordiale. \`E l'oggetto della \lq fotografia\rq ~fatta dal satellite Planck nel 2013 (Figura \ref{cmb}).

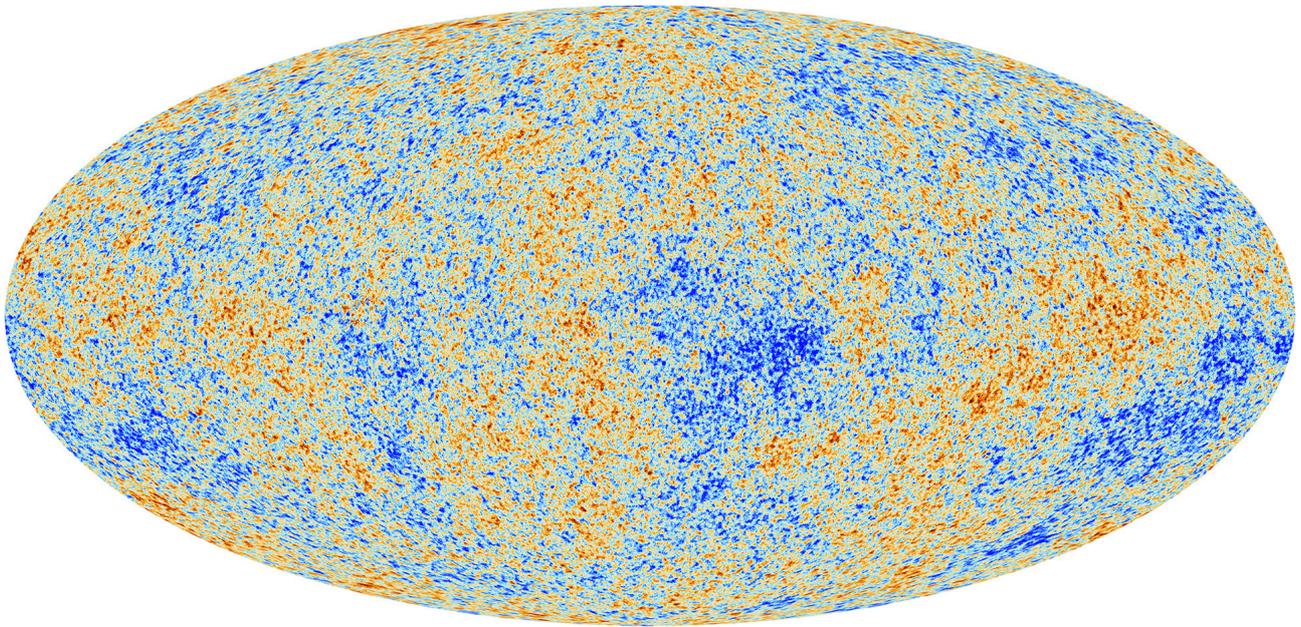
\WideFigure{Planck_CMB}{\label{cmb} La prima foto dell'universo bambino: la radiazione di fondo cosmica. I diversi colori corrispondono a piccolissime variazioni della temperatura ( $\approx 3K$) osservata. Possiamo leggerci l'impronta della  gravit\`a quantistica?}

Le sue variazioni di temperatura, minime, sono prodotte da piccole fluttuazioni dei campi di materia e di interazione (incluso quello gravitazionale) nelle primissime fasi dopo il big bang, e sono queste stesse fluttuazioni, secondo le teorie cosmologiche moderne, ad aver originato le strutture cosmiche (galassie ecc.). L'origine e la dinamica di queste fluttuazioni sono l'oggetto principale di attenzione dei modelli cosmologici, le cui predizioni possono poi essere confrontate con i dati osservativi sulla CMB.

E di modelli cosmologici ne esistono diversi. L'inflazione (a sua volta codificata in un numero di modelli, diversi nei dettagli) postula che l'universo abbia avuto una fase di espansione accelerata subito dopo il Big Bang, e che questa espansione sia dovuta ad un nuovo campo di materia, il cosiddetto \lq inflatone\rq. Le predizioni dettagliate sulla CMB, per\`o, dipendono anche da specifiche ipotesi sullo stato iniziale dell'universo al momento dell'inizio di questa espansione accelerata e sulla dinamica dell'inflatone, giustificabili pienamente solo da una teoria pi\`u fondamentale come la gravit\`a quantistica. Modelli alternativi predicono che il big bang sia in realt\`a parte di un \lq Big Bounce\rq ~cosmico, cio\`e di una dinamica dell'universo che include un periodo di contrazione, alla fine della quale raggiunge un volume minimo e una densit\`a massima (in genere immaginata dell'ordine della scala di Planck), e infine un \lq rimbalzo\rq , un Big Bounce appunto, seguito da un periodo di espansione che \`e quella che osserviamo attualmente. Le fluttuazioni all'origine della CMB verrebbero generate nella fase di contrazione. Ma la natura precisa del bounce e la sua esistenza stessa possono essere giustificati di nuovo solo da una teoria di gravit\`a quantistica. Infine, altri modelli cosmologici contemplano una fase \lq pre-Big Bang\rq ~in cui l'universo \`e statico, cio\`e non evolve e si mantiene a volume costante, seguita da una transizione repentina ad una fase di espansione, quella in cui ci troviamo. Anche questi modelli (detti di \lq universo emergente\rq) possono spiegare le caratteristiche osservate nella CMB. Ma anche questi modelli hanno bisogno di una teoria pi\`u fondamentale che descriva la transizione di fase cosmologica che li caratterizza. Insomma, in tutti questi scenari cosmologici possibili, in diverso modo, la gravit\`a quantistica gioca un ruolo e pu\`o avere conseguenze rilevanti dal punto di vista osservativo.

L'ultima ragione per non disperare sulla possibilit\`a di porre in contatto la gravit\`a quantistica con gli esperimenti, a causa della distanza tra la scala di Planck e i fenomeni comunemente osservati, \`e che la definizione stessa della scala di Planck e l'idea che  solo questa sia quella rilevante per la fisica della gravit\`a quantistica, si basano sulla fisica che conosciamo. Detta cos\`i sembra una banalit\`a. Non \`e sempre cos\`i che procede la scienza? Ci si basa su quello che si conosce per andare oltre. Certo. Il problema \`e che ci aspettiamo, dalla gravit\`a quantistica stessa, cambiamenti drastici sia della Relativit\`a Generale sia della Teoria Quantistica dei Campi, nei loro principi pi\`u fondamentali. E quindi non sappiamo se le deduzioni che facciamo, sulla base di questi principi fondamentali potenzialmente  in via di dismissione, siano cos\`i affidabili. Questo \`e un invito a mantenersi aperti alle sorprese. 

\section{Cosa deve essere la gravit\`a quantistica?}
Chiarito che si tratta di un problema fondamentale e con importanti conseguenze fisiche, come risolvere il problema della gravit\`a quantistica? Come costruire questa teoria? La prospettiva tradizionale, seguita in tutto il secolo scorso, \`e quella pi\`u immediata e naturale. Abbiamo una ottima teoria classica della gravit\`a e dello spaziotempo, la Relativit\`a Generale, e ce ne serve una versione quantistica. Abbiamo a nostra disposizione molte procedure diverse per costruire una teoria quantistica a partire dalla sua formulazione classica; applichiamole alla Relativit\`a Generale. In questa prospettiva, il problema \`e puramente tecnico, e ben definito, almeno per quanto riguarda la costruzione della teoria. Rimane un problema formidabile, dato che le difficolt\`a matematiche nel portare a termine le varie procedure di quantizzazione della Relativit\`a Generale sono molte ed enormi. Non facciamo qui neanche una rassegna breve per descrivere i vari filoni di ricerca basati su questa strategia, che corrispondono pi\`u o meno alle diverse tecniche di quantizzazione utilizzabili. Esistono approcci canonici, covarianti, perturbativi, non-perturbativi \cite{carlip, daniele-book}. Da ognuno abbiamo imparato molto, tutti sono incompleti. Quanto promettenti, \`e giudicato diversamente da ciascun ricercatore, e non \`e cos\`i importante discuterne qui.

Ci\`o che \`e importante \`e dare una idea di quanto anche questa prospettiva conservatrice, che cerca di non introdurre ipotesi radicali o nuove entit\`a fondamentali, si confronta con questioni fisiche e concettuali profonde e difficili.

Tutti questi approcci condividono l'oggetto di base: un campo gravitazionale quantistico, cio\`e uno spaziotempo quantistico (Figura \ref{QuantumSpacetime}), dato che il campo gravitazionale coincide con la geometria dello spaziotempo. Vediamo cosa implica.

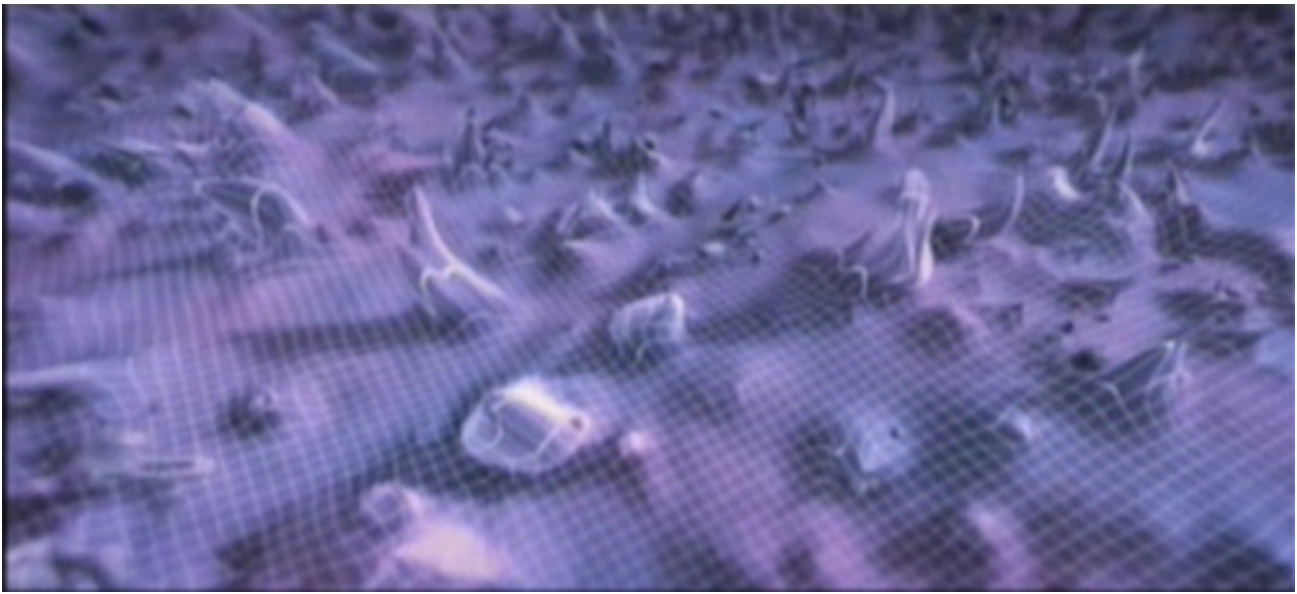
\WideFigure{QuantumSpacetime}{\label{QuantumSpacetime}Uno spaziotempo (continuo, relativistico, geometrico) soggetto a fluttuazioni quantistiche?  (© 1998 Cetin BAL)}

Se lo spazio, il tempo, la geometria sono quantistici, sono necessariamente soggetti a fluttuazioni e determinabili solo in maniera probabilistica. \`E gi\`a difficile avere una intuizione di questo comportamento in Meccanica Quantistica ordinaria, riguardo particelle di materia (un elettrone, ad esempio). Pensate cosa vuol dire avere quantit\`a geometriche, come l'area di un tavolo, il volume di una stanza, la lunghezza di una strada, soggette a fluttuazioni quantistiche (ci aspettiamo che queste fluttuazioni siano quasi assenti per oggetti macroscopici, ma la questione concettuale rimane). Alla base di queste fluttuazioni c'\`e la sovrapposizione di stati quantistici. Nel caso della geometria vuol dire che qualunque affermazione del tipo: \lq\lq La distanza tra l'oggetto A e l'oggetto B \`e X centimetri\rq\rq ~deve essere riformulata come \lq\lq La distanza tra l'oggetto A e l'oggetto B \`e $X_1$ cm con probabilit\'a $P_1$, $X_2$ cm con probabilit\'a $P_2$, ecc.\rq\rq ~con una probabilit\`a non nulla (in generale) che sia qualunque cosa! Lo stesso per affermazioni sugli intervalli temporali tra due eventi, o il volume occupato da un corpo, o la curvatura attorno da un altro. Le quantit\'a geometriche stesse non hanno un valore univoco.
Se questo non vi basta, considerate il fatto che le possibili relazioni causali tra eventi diversi dipendono strettamente dalle loro relazioni geometriche. Tecnicamente, \`e la geometria dello spaziotempo che determina il \lq cono di luce\rq ~di un evento, e distingue tra tutti gli altri eventi quelli che giacciono nel suo futuro (e possono essere da questo influenzati) e quelli nel suo passato (che possono averlo influenzato) (Figura \ref{lightcone}).

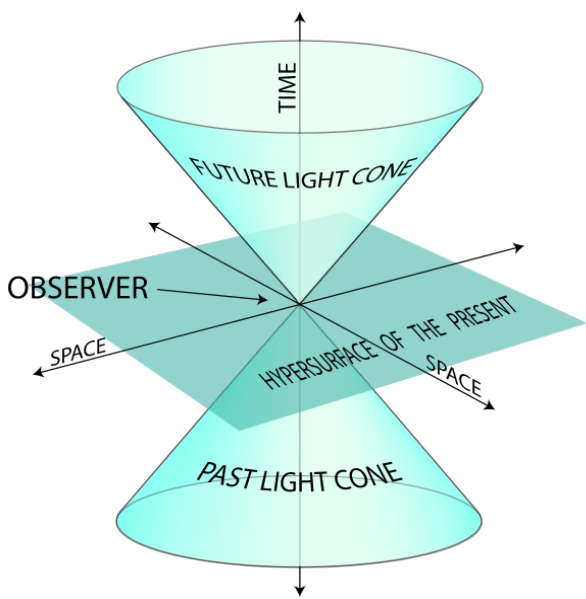
\ColumnFigure{lightcone}{\label{lightcone}Il cono di luce di un osservatore, le limitazioni alle relazioni di causalit\`a, e la distinzione tra passato e futuro. Cosa cambia quando tutto questo diventa quantistico? (da Wikipedia)}

In altre parole, l'affermazione  \lq\lq l'evento A \`e nel passato (futuro) dell'evento B, e pu\`a averlo influenzato (influenzarlo)\rq\rq ~\`e una affermazione sul campo gravitazionale, cio\`e sullo spaziotempo. Se questo ha natura quantistica, anche le relazioni causali, e la distinzione tra passato e futuro, sono soggette a fluttuazioni quantistiche e al principio di sovrapposizione.
E ancora, in tutti i sistemi quantistici che conosciamo, il risultato del processo di quantizzazione a partire dalla teoria classica \`e (anche) che alcune quanti\`a fisiche, che avevano natura continua (potevano prendere qualunque valore intermedio in un dato intervallo, anche infinito) diventano \textit{discrete}. 
Esiste quindi una \textit{risoluzione minima} nella loro misurazione. Pensate cosa pu\`o voler dire nel caso di quantit\`a geometriche: una lunghezza minima, un volume minimo, e quindi una curvatura massima, una energia massima? Non esisterebbe nulla di pi\`u piccolo della lunghezza di Planck, ad esempio, o di pi\`u energetico della energia di Planck. Cosa rimane della nostra intuizione dello spaziotempo come un continuum di eventi, in questo caso? Soltanto la natura continua dei campi che abbiamo quantizzato, che rimarrebbero le entit\'a fisiche fondamentali. Ma questa non si tradurrebbe pi\`u in propriet\`a osservabili anch'esse necessariamente continue, e la loro comprensione andrebbe rivista in profondit\`a.

Al di l\`a di come gli approcci specifici implementano questi aspetti in dettaglio, siamo in presenza di una rivoluzione dei nostri concetti di spazio e tempo e, di conseguenza, della nostra immagine del mondo fin dalle sue fondamenta.

\section{Lo spaziotempo emergente}
Per quanto la prospettiva descritta finora (basata sulla quantizzazione diretta del campo gravitazionale) sia radicale per implicazioni concettuali e fisiche, la prospettiva moderna sul problema della  gravit\`a quantistica lo \`e ancora di pi\`u \cite{daniele-book}.

Vediamo cosa porta i fisici verso questa nuova prospettiva. 

Le \textit{singolarit\`a gravitazionali}, cio\`e le situazioni in cui, secondo la Relativit\`a Generale, la curvatura della spaziotempo cresce senza limiti, come all'interno dei buchi neri o al big bang, sono situazioni in cui la Relativit\`a Generale smette di essere applicabile. Questo fatto indica di sicuro che modifiche quantistiche sono necessarie. Ma molti fisici lo interpretano come un segnale di \textit{inapplicabilit\`a pi\`u generale del continuum spaziotemporale} e dell'idea di campi di interazione (incluso il campo gravitazionale, quantizzato o meno) anch'essi continui che vivono su di esso. Sarebbe cio\`e l'idea stessa di spaziotempo e di campi a venir meno a livello pi\`u fondamentale.

Vari argomenti basati sulla fisica semi-classica, che cercano di stimare gli effetti gravitazionali del comportamento dei campi quantistici, suggeriscono invece che vi sia un limite a quanto precisamente possiamo localizzare gli eventi nel tempo e nello spazio. Questi stessi argomenti portano ad ipotizzare l'esistenza di una \textit{lunghezza minima}, cio\`e di una discretizzazione fondamentale dello spaziotempo. Il risultato \`e di nuovo una dissoluzione del continuum spaziotemporale su cui si basa la teoria dei campi classica e quantistica. Questo avrebbe quindi natura soltanto approssimata, emergente, basandosi proprio sulla  nozione di localit\`a delle interazioni fisiche.

Una natura fondamentalmente discreta del mondo \`e anche ci\`o che indicano con forza i risultati sulla \textit{termodinamica dei buchi neri}, in particolare il loro possedere una entropia finita (sostanziata anche dalla radiazione di Hawking, che essi emettono secondo la fisica semi-classica). Se interpretata alla Bolzmann, infatti, questa entropia misura il numero di gradi di libert\`a (o \lq costituenti elementari\rq) discreti che li costituiscono. Ma i buchi neri non sono altro che particolari regioni di spazio (bench\'e con caratteristiche molto peculiari), quindi stiamo parlando di \textit{costituenti elementari discreti dello spazio(tempo) in  s\`e}!

Le propriet\`a termodinamiche dei buchi neri hanno anche ispirato una quantit\`a di altre ricerche che hanno mostrato come le equazioni della Relativit\`a Generale ammettano una interpretazione termodinamica esse stesse. Possono essere ottenute come equazioni di stato macroscopiche che legano energia ed entropia di gradi di libert\`a microscopici sconosciuti, ma collettivamente caratterizzabili in termini di geometria (campo gravitazionale) e campi di materia.

Insomma, qual \`e l'idea generale suggerita da questi risultati? Che esistono delle \textit{entit\`a microscopiche fondamentali},  e discrete, che 
\textit{costituiscono ci\`o che chiamiamo spaziotempo} e di cui la geometria e i campi con cui lo descriviamo sono soltanto manifestazioni (approssimate) collettive \cite{daniele1}.

Aggiungiamo un altro elemento a supporto di questa idea generale.

Evidenze dell'esistenza di queste entit\`a fondamentali, e suggerimenti concreti sulla loro natura, arrivano direttamente da vari approcci moderni alla gravit\`a quantistica.

In gravit\`a quantistica a loop e nelle teorie di campo su gruppi, le entit\`a fondamentali sono descritte in termini di \textit{spin networks}, cio\`e grafi (o reti, strutture combinatorie fatte di nodi e link (connessioni) tra questi) decorati con ulteriori variabili dinamiche corrispondenti a rappresentazioni di gruppi di simmetria (eg il gruppo di Lorentz) (Figura \ref{spinnet}). Quindi strutture puramente combinatorie-algebriche, che al massimo possono essere messe in corrispondenza con reticoli dotati di qualche forma di geometria discreta, rimpiazzano totalmente variet\`a differenziali, geometria e campi continui.

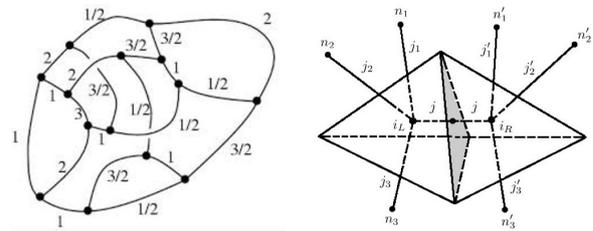
\ColumnFigure{spinnet-tetra}{\label{spinnet}Una spin network: un grafo decorato con semi-interi (rappresentazioni di SU(2) - spins); e la corrispondenza intuitiva con poliedri (qui tetraedri) incollati a formare reticoli estesi.}

Strutture analoghe, corrispondenti a reticoli discreti, sono alla base di tutti gli approcci di gravit\`a quantistica simpliciale (calcolo di Regge, triangolazioni dinamiche) (Figura \ref{triangulation}).

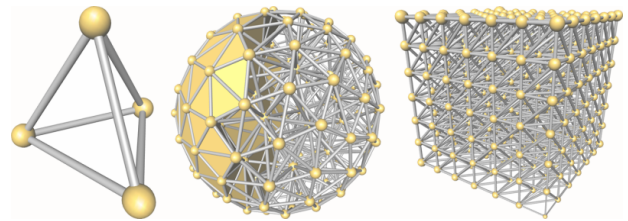
\ColumnFigure{triangulation3}{\label{triangulation} Reticoli simpliciali ottenuti dalla composizione di simplessi 3d (tetraedri).}

La teoria dei \textit{causal sets} \`e similmente basata su entit\`a discrete (reti di relazioni causali elementari) e lontane da quelle alla base delle nozioni usuali di spaziotempo.

Tutti questi approcci, quindi, suggeriscono che lo spaziotempo emerge da strutture pi\`u fondamentali e non spaziotemporali in  s\`e.

In questa prospettiva pi\`u moderna, quindi, uno spaziotempo emergente, non soltanto quantistico, \`e l'oggetto della gravit\`a quantistica. Il problema della gravit\`a quantistica prende una nuova forma, e si carica di compiti ulteriori, ponendosi obiettivi ancora pi\`u radicali.

Descriviamo meglio questi nuovi compiti.

Il problema della gravit\`a quantistica diventa duplice \cite{daniele1, daniele3}: a) individuare e descrivere matematicamente le entit\`a quantistiche, non-spaziotemporali e discrete che costituiscono l'universo al livello pi\`u fondamentale, e la loro dinamica; b) mostrare come lo spaziotempo come lo conosciamo e la sua dinamica effettiva in termini di Relativit\`a Generale e teoria quantistica dei campi emergono in una approssimazione e nelle circostanze appropriate, preceduti, magari, da un regime in cui lo spaziotempo e la geometria sono gi\`a \lq emersi\rq ~e manifestano propriet\`a quantistiche (il regime corrispondente al problema della gravit\`a quantistica tradizionale). Questo, a livello formale. A livello fisico,  c'\`e anche il compito ulteriore di ottenere, da questa nuova descrizione, predizioni qualitativamente chiare e quantitativamente precise su nuovi fenomeni o possibili osservazioni che permettano di testarla.

Si cerca uno spaziotempo emergente a partire da entit\`a non spaziotemporali pi\`u fondamentali \cite{daniele3}.

Ma \textit{cosa} emerge, esattamente, di ci\`o che usiamo per definire lo spaziotempo? Di certo, il campo gravitazionale (la metrica, cio\`e la geometria continua) e la variet\`a differenziale che lo supporta (l'insieme continuo dei \lq punti dello spaziotempo\rq). Forse anche la materia, che viene definita e classificata in base alle sue propriet\`a spaziotemporali (come regisce quando spostata e ruotata, localizzata nel tempo e nello spazio , ecc.). In ultima analisi, forse, \textit{tutto}. Tutte le nozioni che usiamo per fare fisica vanno ripensate, in quanto basate sulle nozioni base di spazio e tempo, e forse sono anch'esse solo approssimate, non fondamentali, emergenti: la localit\'a delle interazioni, la simmetria di Lorentz, il principio di relativit\`a, il principio di equivalenza, ecc. .

\textit{Da cosa} dovrebbero emergere? Beh, questo dipende necessariamente da quale approccio specifico alla gravit\`a quantistica consideriamo. Formalismi diversi presentano candidati diversi per le entit\`a fondamentali. Qualche accenno \`e stato gi\`a dato. Un esempio specifico verr\`a discusso un p\`o pi\`u in dettaglio nel seguito.

E \textit{come}, esattamente, dovrebbero emergere lo spaziotempo e tutte queste nozioni che su esso si basano? Anche qui, molto dipende dal contesto specifico. Approcci diversi troveranno conveniente usare tecniche e idee diverse.

Il punto cruciale \`e per\`o generale \cite{daniele3}. Qualunque siano le entit\`a fondamentali, lo spaziotempo continuo e la sua descrizione in termini di campi di materia e interazione pu\`o emergere solo alla fine di un'approssimazione continua che coinvolga un gran numero di queste entit\`a fondamentali e che sia resa possibile dal risultato della loro dinamica collettiva. A livello pi\`u tecnico, questa intuizione a sua volta suggerisce che un ruolo fondamentale debba essere svolto dal gruppo di rinormalizzazione, che \`e esattamente lo strumento che ci permette di ottenere la dinamica effettiva (e approssimata) macroscopica di sistemi quantistici formati da molti corpi microscopici. Andiamo oltre. Se si deve analizzare la dinamica collettiva di un numero grande di entit\`a quantistiche interagenti, ci si deve anche aspettare che il risultato di questa dinamica collettiva non sia unico. Un sistema di questo tipo si pu\`o organizzare, generalmente, in una molteplicit\`a di \textit{fasi} distinte, alle quali corrisponde una fisica molto diversa. Pensate all'acqua allo stato liquido, che \`e solo una delle fasi in cui le molecole (quantistiche, interagenti) che la costituiscono pu\`o organizzarsi, le altre essendo il ghiaccio (fase solida) e il vapore (fase gassosa). Pensate a tutti gli esempi, compresi quelli molto esotici, che ci presenta la fisica dei sistemi di materia condensata classica e quantistica: fasi macroscopiche con propriet\`a osservabili molto diverse, a partire dagli stessi costituenti elementari (tutta o quasi la fisica della materia condensata si basa su elettroni che interagiscono tramite campo elettromagnetico).  Nel caso di quel sistema peculiare che \`e lo spaziotempo, assumendo sia anch'esso costituito da \lq molti corpi quantistici (non spaziotemporali)\rq ci dobbiamo quindi aspettare che questi possano organizzarsi in fasi diverse. Una di queste deve necessariamente essere caratterizzabile in termini spaziotemporali (la nostra, quella in cui ci troviamo), ma dobbiamo aspettarci anche fasi dove le usuali nozioni di spaziotempo e geometria non emergono sotto \textit{nessuna} approssimazione, fasi interamente non spaziotemporali.

\tcbox{L'analogia tra spaziotempo e fluidi (quantistici)}{
\begin{minipage}{.5\textwidth} \center{\textcolor{blue}{FLUIDO/CONDENSATO}} \begin{itemize} \item Entit\`a  fondamentali: atomi di materia. 
\item Dinamica quantistica fondamentale: processi di interazioni tra particelle e atomi 
(ad esempio scattering, creazione/distruzione). 
\item Fasi continue prodotte dalla dinamica collettiva di molte entit\`a fondamentali: gassosa, liquida, solida, ecc. 
\item Approssimazione utile a livello macroscopico nella fase liquida (e, in parte, gassosa): idrodinamica (o campo medio, ecc.). \end{itemize} \vspace{0.15cm} \end{minipage}  
\begin{minipage}{.5\textwidth}\center{\textcolor{blue}{SPAZIOTEMPO}} \begin{itemize} \item Entit\`a fondamentali: atomi di spazio \item dinamica quantistica fondamentale: interazioni tra atomi di spazio, codificate in strutture discrete 
(ad esempio reticoli). \item Fasi continue prodotte dalla dinamica collettiva di molte entit\`a fondamentali: geometrica/spaziotemporale, altre? \item Approssimazione utile nella fase spaziotemporale/geometrica: in termini di spaziotempo continuo, campi relativistici (classici e quantistici), cosmologia. \end{itemize} \end{minipage}
}

Per finire, chiediamoci di nuovo: dove pu\`o essere la fisica, in tutto  ci\`o? che conseguenze osservative dobbiamo aspettarci?

A questo punto, dovrebbe essere chiaro quello che affermavamo gi\`a in precedenza: aspettiamoci sorprese, in ogni direzione! \textit{Tutte} le strutture e i concetti su cui si basa \textit{tutta} la fisica moderna vengono messi in discussione. Potremmo ben scoprire che gli effetti di gravit\`a quantistica sono ovunque attorno a noi, che magari li avevamo gi\`a sotto gli occhi, ma non li avevamo riconosciuti in quanto tali. Potremmo scoprire che osservazioni e dati sperimentali gi\`a nelle nostre mani, ma che non mettevamo in relazione con la gravit\`a quantistica in quanto si riferivano a fenomeni \lq macroscopici\rq, hanno invece la loro spiegazione nella gravit\`a quantistica, proprio perch\`e lo spaziotempo in s\`e, compresi i suoi aspetti macroscopici, ha natura emergente. Due esempi che sono stati infatti studiati da questo nuovo punto di vista sono la materia oscura e l'energia oscura, fenomeni tuttora in attesa di spiegazione appropriata, e normalmente approcciati da un punto di vista puramente cosmologico, non direttamente collegato a questioni di gravit\`a quantistica.

In alcuni approcci alla gravit\`a quantistica, l'idea di uno spaziotempo come risultato collettivo della dinamica di un sistema a molti corpi \`e preso talmente sul serio da modellizzare letteralmente l'universo come un fluido o un condensato. L'emergere dello spaziotempo continuo viene trattato alla stregua dell'emergere della descrizione idrodinamica del fluido stesso, a partire dalla sua descrizione atomica/molecolare. E la transizione di fase che porta all'esistenza di tale fluido acquista una possibile interpretazione fisica in termini cosmologici, come ci\`o che rende possibile l'esistenza di questo spaziotempo continuo, e identificata con la nascita dell'universo primordiale: il Big Bang.

\section{Teorie di campo su gruppi: il contesto giusto?}
Per illustrare le idee appena discusse, ecco un esempio di formalismo di gravit\`a quantistica basato su entit\`a discrete e non spaziotemporali, in cui lo spaziotempo emerge a livello di dinamica collettiva, e la Relativit\`a Generale ne diventa la descrizione appropriata in un regime idrodinamico. Questo formalismo \`e chiamato \textit{teoria di campo su gruppi} (Group Field Theory, GFT).

Cosa sono queste nuove teorie di campo, intanto? Sono una descrizione \lq atomica\rq e quantistica \textit{dello spaziotempo stesso}, una teoria di campo in cui i \lq quanti\rq ~fondamentali sono i costituenti elementari dello spaziotempo stesso, i suoi \lq atomi\rq ~costitutivi. Per definizione quindi non sono definite su nessuno spaziotempo fisico, ma su spazi in qualche modo ausiliari e pi\`u astratti, corrispondenti a possibili gruppi di simmetria dello spaziotempo che essi dovrebbero generare. E, hanno l'onere di spiegare in che modo lo spaziotempo emerge a partire dai loro quanti fondamentali e dalla loro dinamica quantistica, come in tutti i formalismi di gravit\`a quantistica basati sull'idea di spaziotempo emergente.

I \lq pezzettini di spazio\rq ~descritti dalle GFT sono visualizzabili come poliedri 3d astratti, normalmente tetraedri, ad ognuno dei quali pu\`o essere attribuito un volume, una lunghezza dei lati, una \lq forma\rq ~(schiacciata, equilatera, ecc.) (Figura \ref{GFTquanta}). Uno stato quantistico generico di GFT sar\`a quindi dato da un numero arbitrario di tetraedri ognuno con forma arbitraria; o meglio, essendo una teoria quantistica, sar\`a dato da una sovrapposizione di stati siffatti. Stiamo dicendo davvero che il mondo \`e fatto, l\`i gi\`u alla scala di Planck, da piccoli tetraedrini che si muovono? Si e no. Si, nel senso che stiamo ipotizzando, qui, che una teoria di GFT sia la descrizione corretta del mondo. No, perch\'e la visualizzazione dei gradi di libert\`a fondamentali in termini di tetraedri \`e una guida alla scelta degli ingredienti matematici da includere nei modelli di GFT, per poter estrarre da questi, alla fine di un processo complesso, una geometria continua e uno spaziotempo realistico, ed \`e un aiuto alla visualizzazione di questi modelli e di questo processo. Ma cos\`i come un elettrone non \`e una sferetta puntiforme che gira, anche se ci \`e utile visualizzarlo cos\`i, i quanti di GFT non sono piccole piramidine che si muovono. La ragione pi\`u profonda per cui questa visualizzazione non va presa alla lettera \`e che essa fa uso inevitabilmente di uno spazio ambiente: i tetraedrini, appunto, si trovano e si muovono, nella nostra immagine mentale, in un qualche spazio. I quanti GFT, invece, \textit{sono} lo spazio, o quantomeno lo formano in un qualche regime della loro dinamica collettiva.

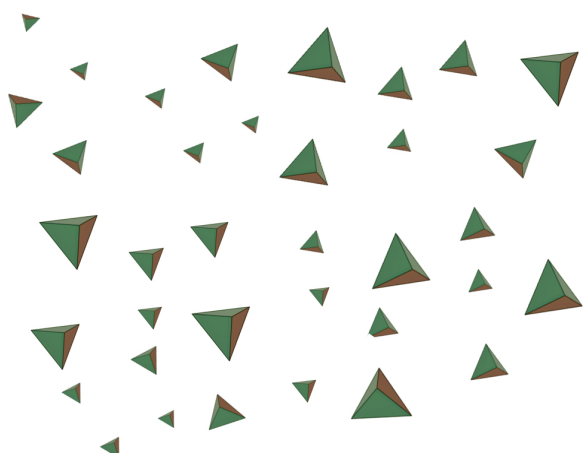
\ColumnFigure{GFTquanta}{\label{GFTquanta} I quanti di GFT: costituenti elementari dello spazio stesso?}

Oltretutto, se possiamo assegnare una geometria discreta, ancorch\'e quantistica, ai singoli quanti di GFT in quanto tetraedri (e parlare delle aree delle loro facce o del loro volume), uno stato generico di GFT, fatto da molti di questi tetraedri, non ha nessuna geometria chiara neanche a livello discreto. I tetraedri che lo compongono non saranno in genere incollati l'uno all'altro a formare alcuna struttura estesa, e, se lo sono, non necessariamente sar\`a possibile visualizzarli come un reticolo geometricamente ben formato.

Insomma, anche prendendo seriamente l'immagine proto-geometrica dei quanti di GFT come tetraedri, siamo ancora ben lontani da uno spaziotempo e una geometria continui come quelli alla base della fisica macroscopica.

Tutto quello che vale per i quanti di GFT, vale anche per i loro processi di interazione, la loro dinamica nel regime in cui un numero limitato di essi viene fatto interagire. I loro processi di interazione, infatti, possono essere messi in corrispondenza anch'essi soltanto, e non sempre, con reticoli, interpretabili come \lq spazitempi\rq \,  discreti (con le stesse limitazioni appena menzionate).

Non molto di pi\`u pu\`o essere ottenuto a questo livello di descrizione. Per andare oltre, alla ricerca di una approssimazione continua soddisfacente, bisogna lavorare con numeri sempre pi\`u grandi di quanti fondamentali, cos\`i come per ottenere una fisica effetiva continua a partire dalla fisica atomica o molecolare bisogna considerare numeri grandi di atomi o molecole interagenti (Figura \ref{LQGspacetime}). In questo formalismo, lo spaziotempo continuo e la sua fisica emergono a partire da queste entit\`a elementari discrete, e la dinamica relativistica del mondo rimane una approssimazione macroscopica del pullulare sottostante dei loro processi quantistici di interazione.

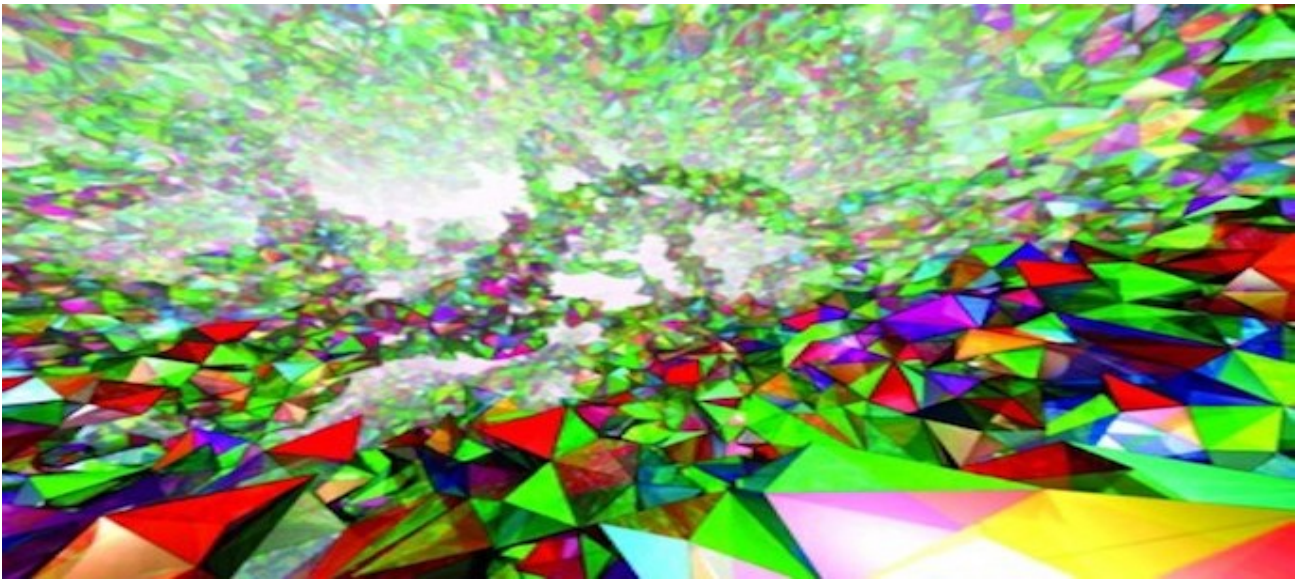
\WideFigure{LQGspacetime}{\label{LQGspacetime} Come emerge lo spazio(tempo) dalla dinamica collettiva di (molti) quanti di GFT?}

Prima di presentare risultati recenti in questa direzione, chiariamo come questo formalismo sia strettamente collegato ad altri approcci di gravit\`a quantistica gi\`a menzionati. Le strutture alla base della gravit\`a quantistica a loop, le spin networks, sono comuni alle GFT, dato che rappresentano una riscrittura equivalente degli insiemi di tetraedri incollati di cui dicevamo. I reticoli alla base degli approcci di gravit\`a quantistica simpliciale sono gli stessi che vengono generati come processi di interazione elementari tra i quanti di GFT. Molte altre tecniche e strutture in comune possono essere evidenziate, cos\`i come inevitabili differenze di prospettive e di  procedure.

Rispetto a questi altri formalismi, le GFT hanno un vantaggio chiave. La riformulazione delle stesse strutture matematiche, che descrivono pressoch\'e le stesse entit\`a fisiche discrete (e non spaziotemporali), nel linguaggio della teoria quantistica dei campi permette di trattarle in maniera potenzialmente molto pi\`u efficiente. Infatti, la riformulazione in termini di teorie quantistiche dei campi \`e esattamente ci\`o che permette di trattare in maniera efficiente i sistemi quantistici a molti corpi, nell'ambito dei sistemi di materia condensata, e di estrarre la loro dinamica effettiva macroscopica (in particolare, nel caso dei fluidi e condensati quantistici, la loro descrizione idrodinamica).
In GFT, ci si propone di fare lo stesso, ma per gli atomi di spazio!

\section{L'emergere dello spaziotempo in GFT: risultati recenti}
Mostrare l'emergere dello spaziotempo in una teoria di gravit\`a quantistica basata su entit\`a non direttamente spaziotemporali o geometriche richiede due cose. Primo, l'esistenza di una fase continua che possa essere descritta in termini di geometria e campi di materia/interazione. Secondo, una riscrittura della dinamica effettiva delle entit\`a fondamentali, in questa fase, che corrisponda alla Relativit\`a Generale (o ad una sua modifica compatibile con le osservazioni), in una approssimazione classica.

A che punto siamo, nelle teorie di campo su gruppi, rispetto a questi punti? Il progresso \`e stato rapido e sostanziale, negli ultimi 10 anni circa, facilitato dal fatto che, come abbiamo anticipato, questi modelli permettono l'applicazione diretta di idee e tecniche standard per i sistemi quantistici a molti corpi (quelle definiti cio\`e \textit{nello} spaziotempo). Riassumiamo alcuni risultati.

Per identificare una fase continua geometrica di un sistema quantistico a molti corpi o di una teoria di campo, bisogna tanto per cominciare avere una mappa chiara delle sue fasi continue. La maniera principe per mappare lo spazio delle fasi di tali sistemi \`e studiarne il cosiddetto flusso di rinormalizzazione. In sintesi estrema, questo vuol dire controllare come cambia la dinamica effettiva del sistema quando si prendono in considerazione interazioni tra numeri sempre maggiori dei suoi costituenti. Quando questi sono stati tutti considerati, e sono in numero infinito, si parla di limite continuo. Le possibili fasi continue sono identificate dai valori che in esse prendono i parametri che caratterizzano le possibili interazioni dei costituenti del sistema. In fasi diverse, la dinamica effettiva pu\`o essere \textit{radicalmente} diversa, e il sistema stesso prendere aspetti cos\`i radicalmente diversi da sembrare un sistema fisico differente. Di nuovo, pensate all'acqua in forma liquida, solida o gassosa, ma le differenze fisiche tra fasi diverse possono essere ancora maggiori, come per esempio in sistemi che sono ottimi conduttori elettrici in una fase, e ottimi isolanti in un'altra.

Lo studio del flusso di rinormalizzazione delle GFT \`e diventato un ambito di ricerca molto attivo, con molti risultati interessanti \cite{sylvain}. Potete immaginare le difficolt\`a tecniche, per\`o! La maggior parte dei risultati pi\`u solidi, infatti, sono stati ottenuti per modelli semplificati, che non posseggono cio\`e tutte quelle caratteristiche che vorremmo in modelli realistici di gravit\`a quantistica. Questo \`e ovviamente un male. La nota positiva, d'altro canto, \`e che i risultati ottenuti finora danno indicazioni che sembrano avere validit\`a piuttosto generale. Per cominciare, i modelli di GFT sembrano consistenti e ben definiti sia quando si considerano interazioni tra pochi costituenti sia quando se ne prendono in considerazione un numero arbitrariamente alto. Questa \`e cosa affatto scontata. Inoltre, sembrano possedere almeno due fasi continue distinte: una apparentemente degenere dal punto di vista della geometria continua, che non sembra avere chances di corrispondere al nostro mondo, e un'altra  pi\`u promettente. Questa corrisponde, in primissima approssimazione, ad una fase \lq condensata\rq , in cui cio\`e i quanti di GFT si organizzano in modo da avere tutti o quasi lo stesso stato quantistico. L'universo diventa un condensato, un fluido quantistico di quanti di GFT.

\`E questa la fase geometrica e pienamente spaziotemporale che corrisponde all'universo che osserviamo? Questi primi risultati sono incoraggianti ma certo non conclusivi. Serve una analisi pi\`u dettagliata sulla possibilit\`a di estrarre da essa la fisica gravitazionale che conosciamo.

Anche su questo secondo aspetto cruciale, i risultati negli ultimi anni sono stati molti e promettenti \cite{daniele2, steffen-lorenzo}. L'attenzione, anche per via delle indicazioni date dal flusso di rinormalizzazione, si \`e concentrata sui condensati di GFT, e sull'estrazione di una dinamica cosmologica. Ma anche in questo caso le indicazioni ottenute sembrano avere validit\`a pi\`u generale.

Il quadro concettuale di riferimento si basa sull'ipotesi che la dinamica effettiva gravitazionale, e in particolare la cosmologia, vadano cercate al livello di approssimazione \textit{idrodinamica} dei modelli di GFT. La giustificazione intuitiva \`e che, in questa approssimazione, ci si concentra sulla dinamica collettiva delle entit\`a fondamentali, trascurando le loro fluttuazioni e interazioni microscopiche, cio\`e quello che ci aspettiamo da una teoria macroscopica continua, e sulle loro osservabili globali, che \`e quello che ci aspettiamo corrisponda ad una dinamica cosmologica.

Gli stati condensati di GFT, poi, con tutti gli atomi di spazio nello stesso stato quantistico, sembrano perfettamente adattati a questa interpretazione cosmologica. Intuitivamente, corrispondono al tipo di geometrie continue utilizzate per descrivere l'universo su scale cosmologiche, cio\`e quelle omogenee, con tutti i punti dello spazio caratterizzati dalla stessa geometria locale. A livello matematico, questa corrispondenza pu\`o essere codificata pi\`u precisamente. La forma matematica degli stati di condensato ha un'altra conseguenza importante: essi sono interamente caratterizzati (di nuovo, nell'approssimazione pi\`u semplice) da un'unica funzione (la \lq funzione d'onda del condensato\rq), pur essendo composti da un numero infinito di gradi di libert\`a. Inoltre, i dati da cui questa funzione dipende sono traducibili direttamente nelle variabili che descrivono spazitempi cosmologici. \`e quindi lo stesso tipo di funzione che viene usata in cosmologia quantistica e interpretata come \lq funzione d'onda dell'universo\rq. Tutto torna, fin qui.

Il vantaggio di considerare stati condensati \`e anche un altro. L'estrazione della loro dinamica effettiva, in approssimazione idrodinamica, a partire dalla dinamica microscopica, \`e immediata, almeno nei casi pi\`u semplici. Anche nel caso dei condensati di GFT, infatti, \`e possibile estrarre l'idrodinamica di condensato per ogni dato modello di partenza. E data la natura e interpretazione della funzione d'onda di condensato, le equazioni dell'idrodinamica assumono la forma di una estensione non-lineare delle equazioni della cosmologia quantistica! Queste possono poi essere usate, come in quel contesto, per derivare predizioni cosmologiche, ma qui esse sono direttamente ricavate dalla teoria fondamentale.

Va bene, ma insomma: cosa dice tutto questo riguardo l'evoluzione del cosmo, assumendo che questa idea di spaziotempo emergente e di universo come condensato siano sensate, e che le GFT siano il contesto giusto per realizzarle?

Per una classe abbastanza generale di modelli di GFT realistici, e sotto l'assunzione che l'universo sia non solo omogeneo, ma anche isotropo (cio\`e con tutte le direzioni spaziali equivalenti, un caso ancora pi\`u semplice, ma alla base di molta cosmologia), la dinamica effettiva del volume dell'universo pu\`o essere dedotta esplictamente dalle equazioni idrodinamiche del condensato.

Le evoluzioni possibili sono diverse, ma sono caratterizzate da alcuni punti in comune, tutti piuttosto eccitanti.
Tanto per cominciare, l'universo-condensato si espande nel tempo, e fin qui tutto coincide con quello che osserviamo nell'universo reale. Inoltre, quando l'universo cresce abbastanza, le equazioni dinamiche che lo governano sono ben approssimate da quelle della Relativit\`a Generale, che \`e l'altra condizione necessaria per continuare a fidarsi delle equazioni medesime. Continuando ad seguire l'evoluzione dell'universo-condensato e la sua espansione, i modelli di GFT in cui le interazioni tra i costituenti fondamentali sono forti predicono che l'espansione si fermer\`a e comincer\`a invece una fase di contrazione cosmica, con l'universo che assume volumi sempre pi\`u piccoli. Gli stessi volumi piccoli che aveva all'inizio della sua espansione, vicino al big bang. E prima? e dopo? Cosa succede al big bang? L'universo-condensato, questo dicono le equazioni idrodinamiche, attraversa un \lq big bounce\rq, rimbalza, passando da contrazione ad espansione. Il big bang, con la sua singolarit\`a gravitazionale \`e sostituito da una regione puramente quantistica a volume minimo. Questo pu\`o avvenire infinite volte, in infiniti cicli di espansione-contrazione.

La gravit\`a quantistica ha fatto quindi la sua scelta, tra tutti i modelli di universo primordiale che abbiamo elencato in apertura, scegliendo uno scenario di  Big Bounce?  Non cos\`i in fretta!
Non solo perch\`e la scienza non \`e scienza senza innumeravoli \lq ma\rq, \lq  per\`o\rq e \lq a meno che\rq, cio\`e per la cautela intellettuale e lo scetticismo che ci definiscono (dovrebbero definirci) in quanto scienziati.
Ma per ragioni pi\`u tecniche. Intanto, perch\'e quando le interazioni tra quanti di GFT sono abbastanza forti, 
l'espansione in \lq uscita\rq \,
dal Big Bounce sembra essere di tipo accelerato, come negli scenari inflazionari (ma con un accelerazione generata invece da puri effetti di gravit\`a quantistica, senza bisogno di campi inflatonici aggiuntivi). Quindi, in questi casi siamo in presenza di uno scenario misto Big Bounce-inflazione che va studiato in maggior dettaglio.
Pi\`u importante ancora, tutto ci\`o che abbiamo detto finora si applica non solo alll'interno dell'approssimazione idrodinamica, e quando si guarda soltanto alla dinamica del volume cosmico, senza calcolare cosa succede alle fluttuazioni quantistiche dello stesso, n\`e introdurre altri campi di materia, n\`e studiare altre osservabili geometriche, n\`e considerare cosa succede alle piccole perturbazioni della geometria (che introducono le inomogeneit\`a alla base della radiazione di CMB). Non sappiamo, al momento, come queste complicazioni aggiuntive, sicuramente necessarie, alterano lo scenario appena descritto. Una possibilit\`a in particolare va menzionata. La dinamica delle fluttuazioni e delle perturbazioni attorno alle configurazioni di condensato semplice che sono state considerate finora potrebbero alterare drammaticamente la dinamica vicino al big bounce, e in maniera tale da rendere l'approssimazione idrodinamica totalmente inadeguata. Questo in particolare \`e quello che ci dovremmo aspettare se la nascita dell'universo, la fase subito a ridosso del Big Bang classico, fosse in realt\`a il risultato di una transizione di fase del sistema di quanti di GFT, da una fase non geometrica e non spaziotemporale ad una fase geometrica, in cui i concetti di spazio e tempo possono essere applicati. Questa transizione sarebbe il processo di \textit{condensazione} degli atomi di spazio che d\'a origine alla fase condensata in cui tutti i risultati appena menzionati sono stati ottenuti. In questo caso, lo scenario cosmologico sarebbe quello di \lq universo emergente\rq menzionato in precedenza.

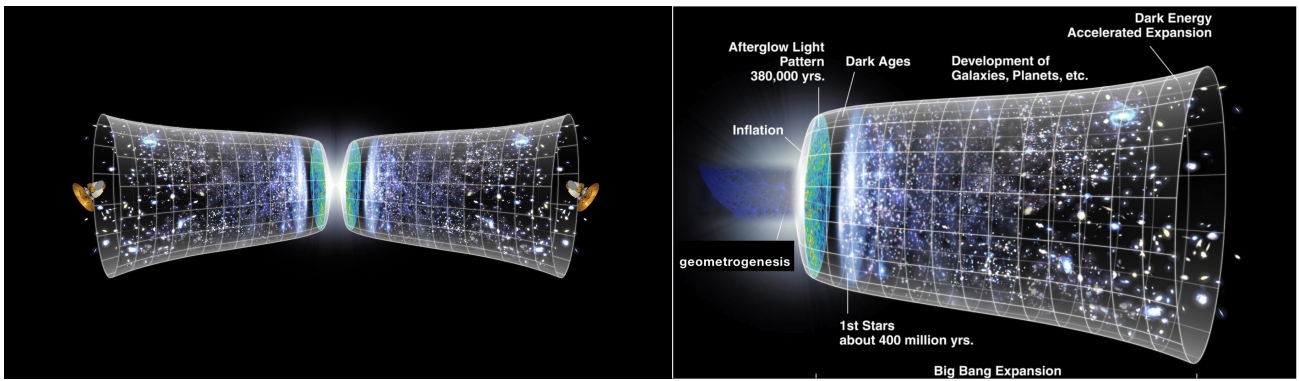
\WideFigure{bounceVSgeometrogenesis}{\label{bounceVSgeometrogenesis} Big Bounce (con un universo classico prima e dopo)? oppure una condensazione cosmica (geometrogenesi) a formare lo spaziotempo,partendo da una fase senza spazio e senza tempo?}

Manca ancora molto lavoro, prima di poter decidere cosa \`e successo all'inizio dell'evoluzione cosmica (Figura\ref{bounceVSgeometrogenesis}). La speranza \`e che questa sia il sentiero concettuale e il contesto formale giusto per rispondere a questa domanda, e per realizzare l'idea di spaziotempo emergente in  gravit\`a quantistica. La speranza ulteriore \`e che quanto stiamo imparando in questo specifico approccio al problema abbia in realt\`a validit\`a pi\`u generale.

Nel frattempo, altri risultati sono stati ottenuti, nell'ambito della cosmologia di GFT. Riguardano la dinamica delle anisotropie, altri aspetti di questa idrodinamica cosmica  e, soprattutto, la teoria delle perturbazioni cosmologiche, la base necessaria per mettere in contatto la gravit\`a quantistica con la cosmologia osservativa. 


\

Per concludere. Le domande della gravit\`a quantistica riguardano le fondamenta stesse della nostra comprensione del mondo fisico. La rivoluzione dei concetti di spazio, tempo e materia che ci si aspetta dalla sua costruzione \`e radicale, e tanto di pi\`u nella prospettiva moderna che vede lo spaziotempo stesso come emergente.
Privati della base concettuale su cui poggia il nostro abituale sguardo sul mondo, cerchiamo nuove fondamenta. Cerchiamo un nuovo pensiero, degli occhi nuovi per ammirare l'universo.


\AuthorsBio{Daniele Oriti}{dirige un gruppo di ricerca in  gravit\`a quantistica al Max Planck Institute for Gravitational Physics di Potsdam, in Germania, dal 2009. Prima, ha fatto ricerca all' Universit\`a di Cambridge, UK, all' Universit\`a di Utrecht, Olanda, e al Perimeter Institute for Theoretical Physics in Canada. Nel 2008 ha ricevuto il Sofja Kovalevskaja Prize dalla A. von Humboldt Foundation. Lavora con diversi formalismi, concentrandosi sulle teorie di campo su gruppi. Si occupa sia degli aspetti matematici che di quelli fisici della  gravit\`a quantistica, con particolare attenzione alla cosmologia, ma anche delle sue implicazioni filosofiche. \`E attualmente membro dell'Arnold 
Sommerfeld Center for Theoretical Physics della Ludwig-Maximilians Universit\"at di Monaco, e assegnatario di un Heisenberg Grant della Deutsche Forschung Gesellschaft.}

\end{document}